\documentclass[twocolumn,aps,pra,showpacs,superscriptaddress,amsfonts,amssymb,amsmath,showkeys]{revtex4-2}

\usepackage{graphicx}% Include figure files
\usepackage{dcolumn}% Align table columns on decimal point
\usepackage{bm}% bold math

\usepackage{graphicx}
\usepackage{appendix}
\usepackage{amsmath}
\usepackage{mathrsfs}
\usepackage{bm}
\usepackage{color}
\usepackage{CJKutf8}

\usepackage{placeins}
\begin{document}

\preprint{APS/123-QED}

\title{Realization of Landau-Zener Rabi Oscillations on optical lattice clock}% Force line breaks with \\
%\thanks{}%

\author{Wei Tan}
 \affiliation{Key Laboratory of Time and Frequency Primary Standards, National Time Service Center, Chinese Academy of Sciences, Xi’an 710600, China}
 \affiliation{Hefei National Laboratory, Hefei 230088, China}
\author{Wei-Xin Liu }
\affiliation{Department of Physics, Xinzhou Normal University, Xinzhou 034000, China}
\author{Ying-Xin Chen }
 \affiliation{Key Laboratory of Time and Frequency Primary Standards, National Time Service Center, Chinese Academy of Sciences, Xi’an 710600, China}
 \affiliation{Hefei National Laboratory, Hefei 230088, China}
\affiliation{School of Astronomy and Space Science, University of Chinese Academy of Sciences, Beijing 100049, China}
\author{Chi-Hua Zhou }
 \affiliation{Key Laboratory of Time and Frequency Primary Standards, National Time Service Center, Chinese Academy of Sciences, Xi’an 710600, China}
\author{Guo-Dong Zhao }
 \affiliation{Key Laboratory of Time and Frequency Primary Standards, National Time Service Center, Chinese Academy of Sciences, Xi’an 710600, China}
 \affiliation{Hefei National Laboratory, Hefei 230088, China}
\affiliation{School of Astronomy and Space Science, University of Chinese Academy of Sciences, Beijing 100049, China}
\author{Hong Chang}
\thanks{corresponding author: changhong@ntsc.ac.cn}
 \affiliation{Key Laboratory of Time and Frequency Primary Standards, National Time Service Center, Chinese Academy of Sciences, Xi’an 710600, China}
 \affiliation{Hefei National Laboratory, Hefei 230088, China}
\affiliation{School of Astronomy and Space Science, University of Chinese Academy of Sciences, Beijing 100049, China}
\author{Tao Wang}
\thanks{corresponding author: tauwaang@cqu.edu.cn}
\affiliation{Department of Physics, and Chongqing Key Laboratory for Strongly Coupled Physics, Chongqing University, Chongqing, 401331, China}
\affiliation{Center of Modern Physics, Institute for Smart City of Chongqing University in Liyang, Liyang, 213300, China}
%\affiliation{%


%\affiliation{
%\affiliation{
 %Third institution, the second for Charlie Author
%\affiliation{%



\begin{abstract}
Manipulating quantum states is at the heart of quantum information processing and quantum metrology. Landau-Zener Rabi oscillation (LZRO), which arises from a quantum two-level system swept repeatedly across the avoided crossing point in the time domain, has been suggested for widespread use in manipulating quantum states. Cold atom is one of the most prominent platforms for quantum computing and precision measurement. However, LZRO has never been observed in cold atoms due to its stringent requirements. By compensating for the linear drift of the clock laser and optimizing experimental parameters, we successfully measured LZRO on the strontium atomic optical clock platform under both fast and slow passage limits within $4$ to $6$ driving periods. Compared to previous results on other platforms, the duration of the plateau is $10^4$ times longer in the optical lattice clock. The experimental data also suggest that destructive Landau-Zener interference can effectively suppress dephasing effects in the optical lattice clock, paving the way for manipulating quantum states against various environmental effects in cold atomic systems.
\end{abstract}

%Use showkeys class option if keyword
\maketitle

%\tableofcontents

\section{\label{sec:level1}Introduction} %\textbackslash\textbackslash
\label{section1}
Cold atom is among the most promising candidates for quantum computing and plays important roles in the precise measurement of acceleration, electric and magnetic fields, time, frequency, and more \cite{ref17,ref18,ref19,ref20,ref21}. Manipulating the quantum states of atoms, typically using Rabi oscillation excited by external laser or microwave fields, is a key step in all these applications. However, due to various dephasing effects, it is challenging to control the state universally in an atomic ensemble. For example, in atoms trapped in a harmonic trap at finite temperatures, the atom-laser coupling strength varies with the motional state quantum number due to coupling with phonons. This variation makes it difficult to apply a perfect $Pi/2$ pulse to all the atoms, thereby degrading the precision of Ramsey-type interferometers.

Landau-Zener-Stuckelberg interference appears when modulating two energy states to traverse the avoided level crossing points periodically \cite{ref1,ref2,ref3,ref4,ref5,ref6,ref7,ref8,ref9,ref10,ref11,ref12,ref13,ref14,ref15,ref16,refn1,refn2,refn3}. If the interference is allowed to occur at multiple consecutive Landau-Zener (LZ) transitions, periodic oscillations of the population in different energy levels will occur, which is similar to the Rabi oscillation \cite{ref24,ref25,ref26}. This kind of time-domain oscillation is usually named Landau-Zener-Rabi oscillation (LZRO). LZROs have been proven to be against various dephasing effects and have been widely used in manipulating quantum states globally for electric spin, quantum dot and NV center \cite{refn4,ref18,ref19, ref27}. Thus, it is natural to generate this method for cold atom systems. However, the realization of LZROs requires strict experimental conditions that the physical system needs a long coherence time for the long sequence of Landau Zener process. At present, it has never been obtained in the cold atomic system.

On the optical clock platform, the transition between the two clock states $\left| 0 \right\rangle $ and $\left| 1 \right\rangle $ belongs to the dipole forbidden transition, and the clock laser used for detection also has a long enough coherent time \cite{ref31,refn5}, which provides a good opportunity for the detection of LZRO. Another point of benefit is that the atoms are in a light trap of magic wavelengths, thus the construction of the Hamiltonian quantity can be very precise \cite{refn6}. The LZ Hamiltonian can be realized on optical clock platform by periodic modulating transition frequency \cite{ref35,ref36,ref37}.

However, the drift of the clock laser could still be a challenge for detecting LZRO, which requires the fixed detuning. In this paper, we compensated the linear drift of the clock laser and chose the system parameter to minimize the effect on LZRO for non-linear drift. An external frequency modulation of time synchronization is added to the frequency of clock laser by an acousto-optic modulator (AOM), which can achieve accurate modulation frequency, amplitude, and phase. Within the above operation, the LZRO is detected on the strontium atomic optical clock platform under the fast and slow passage limit.

Our paper is organized as follows. In Section \ref{section2}, we described the experimental setup and theoretical principle for LZRO. In Section \ref{section3}, we discuss the experimental results.  In Section \ref{section4}, we present our main conclusions.

\par
%On the other hand, LZRO measures the change of excitation rate of fixed-detuning position of clock transition with evolution time, it is necessary to accurately locate the fixed-detuning frequency. It is a simple method to measure the fixed-detuning position accurately by a symmetric spectrum. However, in LZSM interference, the spectrums are asymmetric in many cases, and because of the frequency drift of the clock laser, it is difficult to determine the fixed-detuning position based on such spectrums. Another method is to compensate the linear drift of the clock laser, so that the frequency of the clock light remains approximately constant. By scanning the evolution time during the detection process, fast and accurate LZRO measurements can be realized.
\par
%In this paper, the linear drift of the clock laser is compensated, which can keep the clock laser frequency in a fixed position—zero-detuning within the effective time, so as to achieve fast Rabi spectrum measurement. In order to achieve large modulation amplitude of clock transition, an external frequency modulation of time synchronization is added to the frequency of clock laser by AOM, which can achieve accurate modulation frequency, amplitude and phase. 

\section{Experimental setup and principle}
\label{section2}
\begin{figure}
\centering
\includegraphics[width=0.5\textwidth]{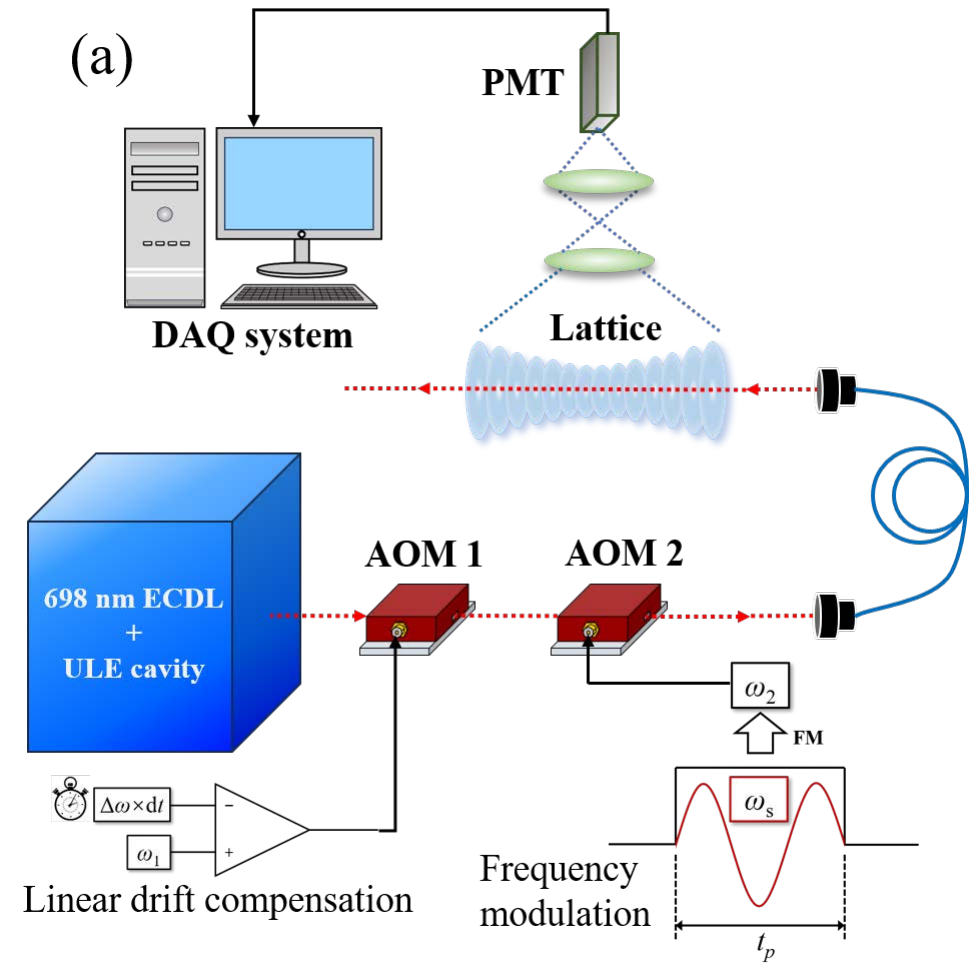}
\includegraphics[width=0.5\textwidth]{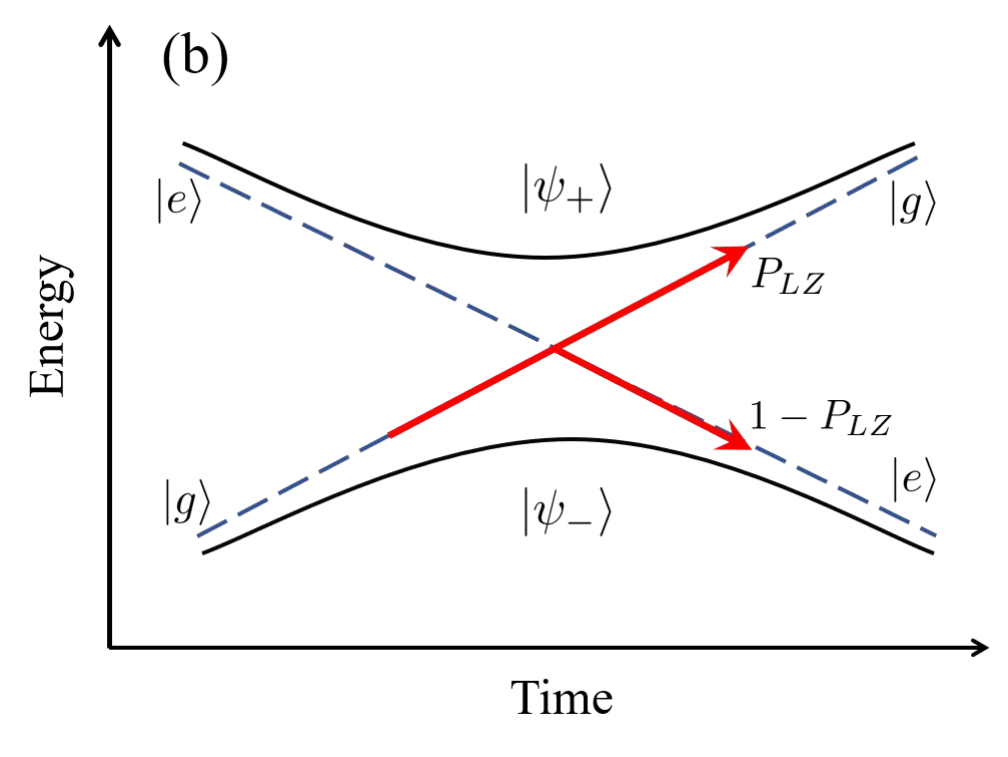}
\caption{(a) Experimental scheme. The polarization directions of the 813.43 nm magic wavelength lattice light and the clock laser are perpendicular to the ground and coincide. The diameter of the 698.45 nm clock laser beam is around 800 ${\rm{\mu }}$m, which is much larger than the lattice beam waist diameter. The 698 nm external cavity diode laser (ECDL) is locked onto a 10 cm ULE glass cavity using Pound-Drever-Hall (PDH) methods. The gate modulation is triggered by the data acquisition (DAQ) system, and its width can be set manually. The lattice imaging lens, composed of two lenses, converges weak lattice signals onto the photomultiplier tube (PMT), and the detected signals are recorded by the data acquisition and processing system. (b) Schematic diagram of LZ transition. The dashed lines represent the diabatic energy levels, while the solid lines represent the adiabatic energy levels. When the system's energy levels undergo an avoided crossing, a non-adiabatic transition occurs between the adiabatic states, as given by Eq.\eqref{3}.}
\label{fig1}
\end{figure}

Optical lattice clock system could be described as an ensemble of two level atoms trapped in identical harmonic traps formed by 'magical wavelength' lattice laser, and an ultra-narrow clock laser excited the atoms from ground state (${}^1{S_0}$) to the long lived state (${}^3{P_0}$) \cite{ref31,refn5}. In our experiments, Approximately $10^{4}$ $^{87}$Sr atoms were trapped in an optical lattice with a trap depth of 90 $Er$, in which $Er$ means recoil energy. The axial and radial temperature of the cold atoms were 4.7 ${\rm{\mu }}$K and 6.3 ${\rm{\mu }}$K, respectively. The longitude modulation of Landau-Zener Hamiltonian was obtained by periodically driving the clock laser frequency. To stabilize the driving phase of 698 nm clock laser, an important factor for the experiments, we implemented the driving by using a burst-mode sinusoidal low-frequency signal delivered to the laser frequency via an AOM. The burst pulse was triggered by the clock sequence. This setup allows convenient and accurate adjustment of parameters such as frequency, amplitude, and initial phase of the modulation waveform. The device schematic diagram is shown in Fig. \ref{fig1}(a).
\par 
%In our experiments, the realization of the frequency modulation of the clock transition is obtained by the modulation of the clock laser frequency, which is achieved by a two-channel arbitrary waveform generator. A gate pulse sinusoidal modulated signal controlled by an external trigger is generated by channel 1, through which the sinusoidal signal frequency is the modulated frequency. The modulated signal is loaded to the carrier frequency of channel 2, and the external frequency modulation function is turned on in channel 2, which can adjust the modulation phase and modulation amplitude. This results in a phase-controlled trigger frequency modulated clock laser.

Under the modulation, the form of the clock laser becomes $\cos \left[ {\int {\left( {{\omega _p} - A{\omega _s}\cos {\omega _s}t} \right)dt} } \right]$, where the $A$ and ${\omega _s}$ are respectively the amplitude and the frequency of the modulation signal, ${\omega _p}$ is the average clock laser frequency. After transforming to a proper rotating frame and applying the rotating-wave approximation, the Hamiltonian for the dynamics of clock states becomes 
\par
\begin{equation}
{\hat H_{\vec n}}\left( t \right) = \frac{h}{2}\left[ {\delta  + A{\omega _s}\cos \left( {{\omega _s}t} \right)} \right]{\hat \sigma _z} + \frac{{h{g_{\vec n}}}}{2}{\hat \sigma _x},\label{1}
\end{equation}
where $h$ is the Planck constant, $\delta  = {\omega _0} - {\omega _p}$ is the clock laser detuning. ${\hat \sigma _{z\left( x \right)}}$ is the Pauli matrix, and
\begin{equation}
 {{g_{\vec n}} = g{L_{{n_z}}}\left( {\eta _z^2} \right){L_{{n_x}}}\left( {\eta _x^2} \right){e^{ - {{\eta _z^2} \mathord{\left/
 {\vphantom {{\eta _z^2} 2}} \right.
 \kern-\nulldelimiterspace} 2}}}{e^{ - {{\eta _x^2} \mathord{\left/
 {\vphantom {{\eta _x^2} 2}} \right.
 \kern-\nulldelimiterspace} 2}}}} \label{2}
\end{equation}
 is the modified coupling strength of clock laser to clock transition at external eigenstates $\vec n = \left( {{n_z},{n_x}} \right)$. ${\eta _z} = 0.25$ and ${\eta _x} = 0.022$ are the longitudinal and transverse Lamb-Dicke parameters, respectively. ${L_n}\left(  \cdot  \right)$ is the $n$th order Laguerre polynomial.
\par
When the longitudinal term of Hamiltonian Eq. \eqref{1} approaches zero, the energy spetrum experience an avoided crossing point, thus the LZ transition occurs, as shown in Fig. \ref{fig1}(b). Assuming the system is initialized in the ground state, the probability of the system still in the ground state after avoided crossing point is given by the so-called LZ formula
\begin{equation}
{{P_{LZ}} = \exp \left( { - \frac{{\pi g_{\vec n}^2}}{{2v}}} \right)} \label{3}
\end{equation}
with $v = A\omega _s^2$, which is the sweep rate of energy levels at the avoided crossing point for the zero-detuning \cite{refn6}. Among LZ transitions between two avoided crossing points, the evolution of the system is assumed to adiabatically follow the instantaneous eigenstates of Hamiltonian Eq. \eqref{1}, thus acquiring different phases for the ground and excited paths. Each LZ transition brings atom waves in the two paths together, leading to interference. Repeatedly sweeping across the avoided crossing point in time zone will induce LZRO between different energy levels. For LZRO, it is convenient to define two bases: the adiabatic basis, which consists of the instantaneous eigenstates and is labeled as $\left| -\right\rangle$ and $\left| +\right\rangle$; and the diabatic basis, which consists of the ground and excited states of atoms and is labeled as $\left| g\right\rangle$ and $\left| e\right\rangle$. These two bases can be transformed into each other. In experiments, the atomic population probability in the diabatic basis can be detected. Based on this, the corresponding population probability in the adiabatic basis can be calculated. 
\par

%In order to realize the fast detection of LZRO, two acoustic optical modulators (AOMs) are used in the experiment to realize the line drift compensation and frequency modulation of the clock laser respectively. The sum of the carrier frequencies of the two AOMs is the difference between the ULE cavity mode frequency and the clock transition frequency. Firstly, by feedback the loop on the optical clock system, the linear drift speed of the clock laser slowly changed between 0.08-0.110 Hz/s. Due to the difference in laboratory temperature, humidity, and other environmental factors every day, the drift speed of the ULE cavity is slightly different. But the frequency shift over a two-hour period basically satisfies the linear relationship. Therefore, applying a fixed period of linear drift compensation to AOM1 can keep the clock laser in the fixed-detuning position of the clock transition spectrum during the experimental time. The LZRO spectrum can be quickly obtained by gradually increasing the clock laser detection time in time sequence. It can also scan the clock laser frequency to realize the detection of LZSM interferences.
%\section{AABBHHV\label{2}}
Detecting LZRO requires fixed detuning $\delta=0$, which is challenging for our system due to the uncontrollable drift of the clock laser. For conventional Rabi oscillation, the zero detuning point is determined from the symmetric Rabi spectrum at a fixed probing time. However, this method fails for detecting LZRO, as the Rabi spectrum is no longer symmetric with $\delta=0$. Therefore, to achieve fast and precise detection of LZRO, it is essential to stabilize the center frequency of the clock laser. Based on our loop-on experiments, we observed a slow variation in the frequency drift rate of the clock laser, ranging from 0.06 to 0.09 Hz/s over one day, see Appendix A for more details.  This drift is caused by variations in the eigenfrequency of the ultra-low expansion (ULE) cavity due to factors such as temperature, humidity, and vacuum conditions. Fortunately, the frequency drift over a two-hour period generally follows a linear trend. Short-term frequency fluctuations (with a stability of less than 5.0E-15 at 1s) mainly arise from vibrations and impose limits on the linewidth of the ULE cavity. By optimizing the Fourier limit of the clock transition line, it is possible to reduce the impact of short-term frequency noise on LZRO. To achieve this, a fixed stepped linear drift compensation is applied to AOM1. This compensation, implemented through a loop-on per-measurement approach, helps maintain a fixed-detuning position for the center frequency of the clock laser throughout the entire duration of the experiment.

\section{Results and discussions}
\label{section3}
For LZRO, two limiting cases named fast-passage limit ($1-P_{\rm LZ}\ll1$) and slow-passage limit ($P_{\rm LZ}\ll1$) are special examples that exhibit distinct oscillation behavior. Specifically, in the fast-passage limit, the system evolves through the diabatic states $\left| g \right\rangle $ and $\left| e \right\rangle $, while in the slow-passage limit, the system follow the adiabatic states $\left| - \right\rangle $ and $\left| + \right\rangle $. When the system sweeps across the avoided crossing points repeatedly, the phase accumulated during the evolution will induce constructive and destructive interference, which is reflected in the oscillation behavior between the two diabatic or adiabatic states \cite{refn1,ref24,ref25}.
\begin{figure}
\includegraphics[width=0.5\textwidth]{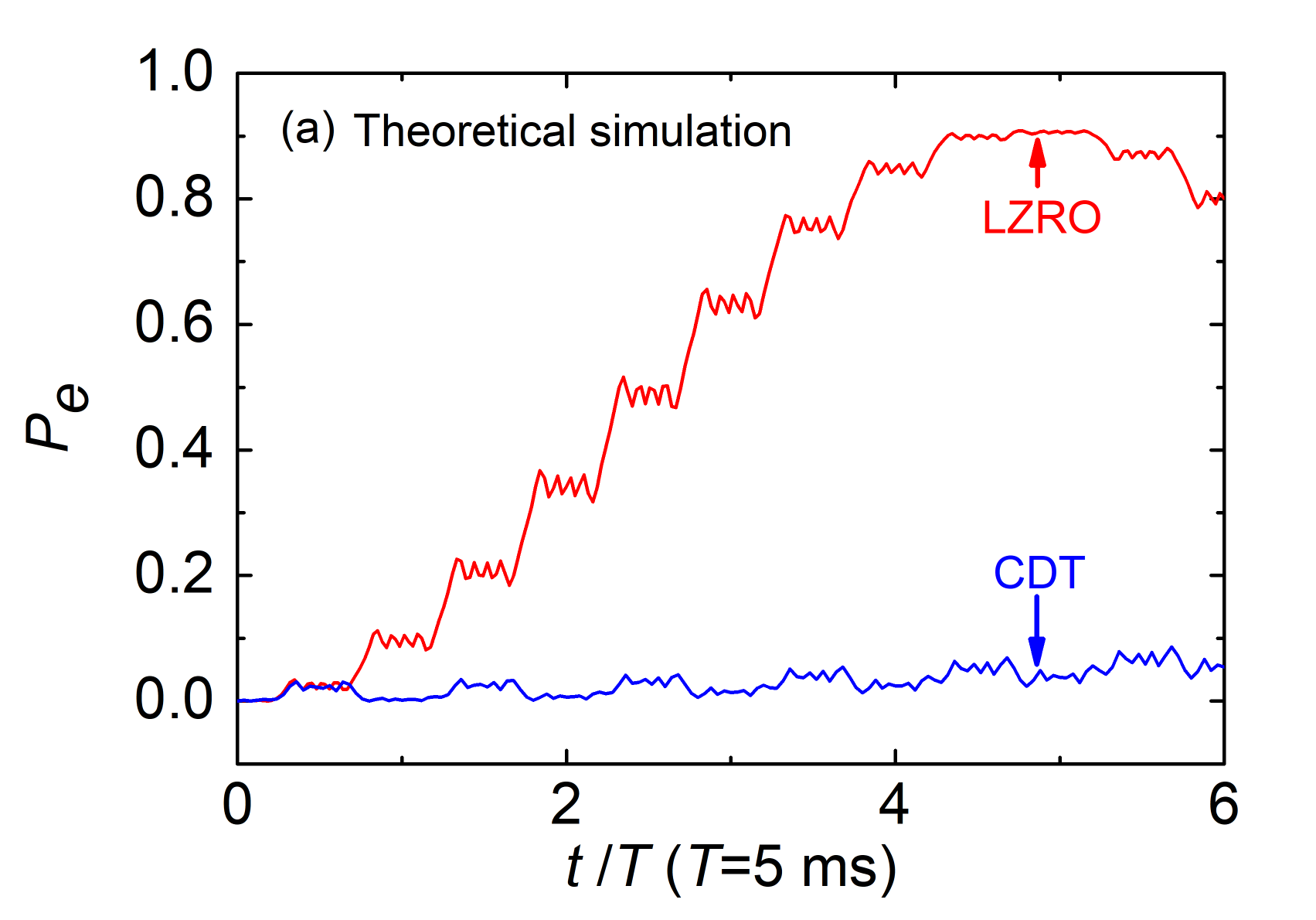}
\includegraphics[width=0.5\textwidth]{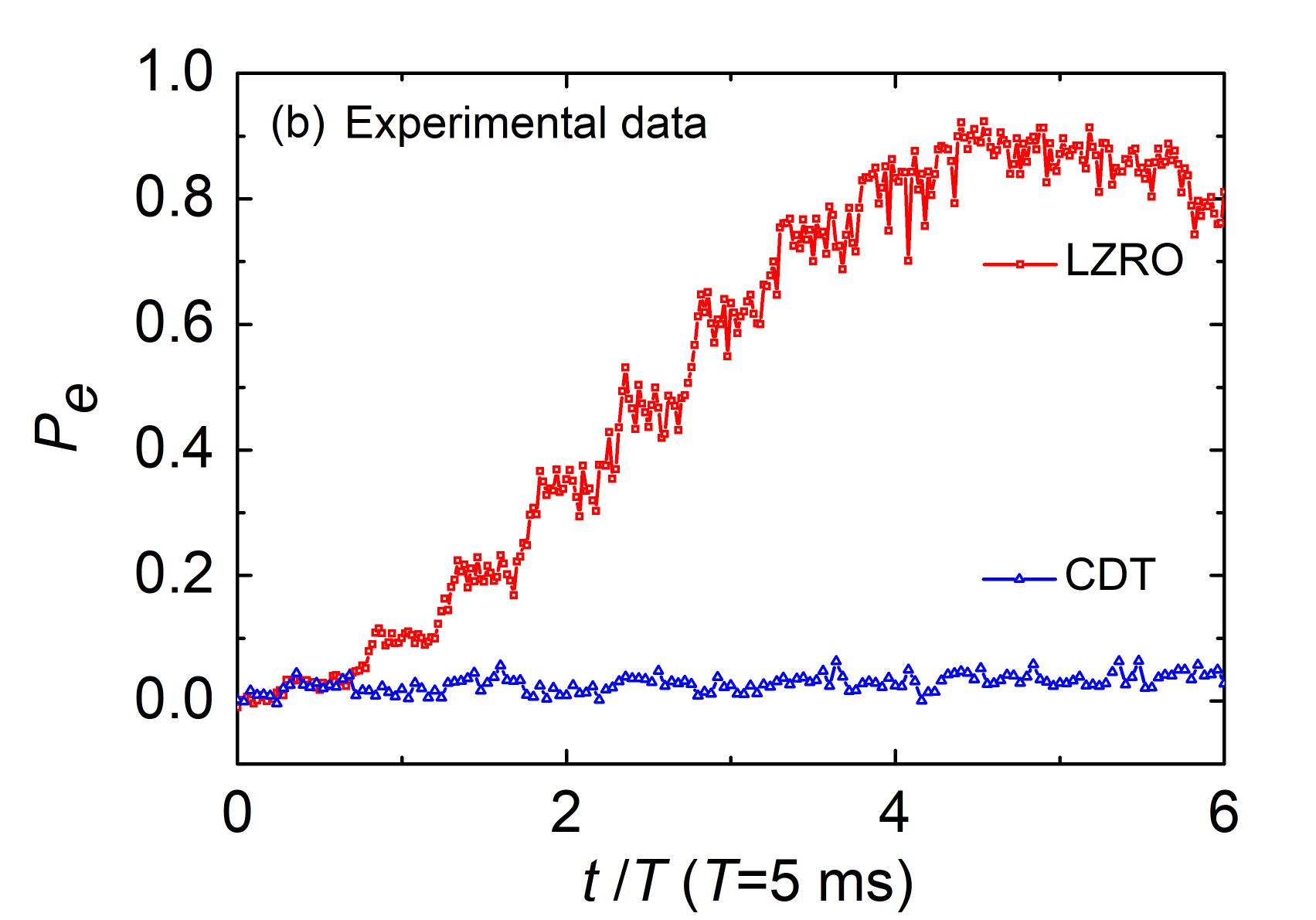}
\caption{The dependence of Excitation rate of the clock transition on the detection time $t$ in the fast-passage limit. Six oscillation periods were performed, each with a period of 5 ms. (a) Theoretical numerical simulation results. The LZRO and CDT spectra are plotted (LZRO in red and CDT in blue). (b) Experimental data. The upper red line represents the constructive LZRO spectrum, while the lower blue line represents the destructive spectrum corresponding to CDT.}
\label{fastp}
\end{figure}
We first measured the constructive and destructive interference induced by a sequence of LZ transitions under fast-passage limit. In this experiment, the parameter $g$ was tuned to near 120 by carefully adjusting the incident 698 $nm$ laser power, and set $ 2\pi g/\omega _s = 0.6$. When the modulation amplitude $A$ was closing to 13.3 and 11.55, giving ${P_{LZ}} \approx 0.97$ (fast-passage regime), the LZ interferences satisfied the conditions for constructive and destructive interference, respectively. 
\par
\begin{figure}
\includegraphics[width=0.5\textwidth]{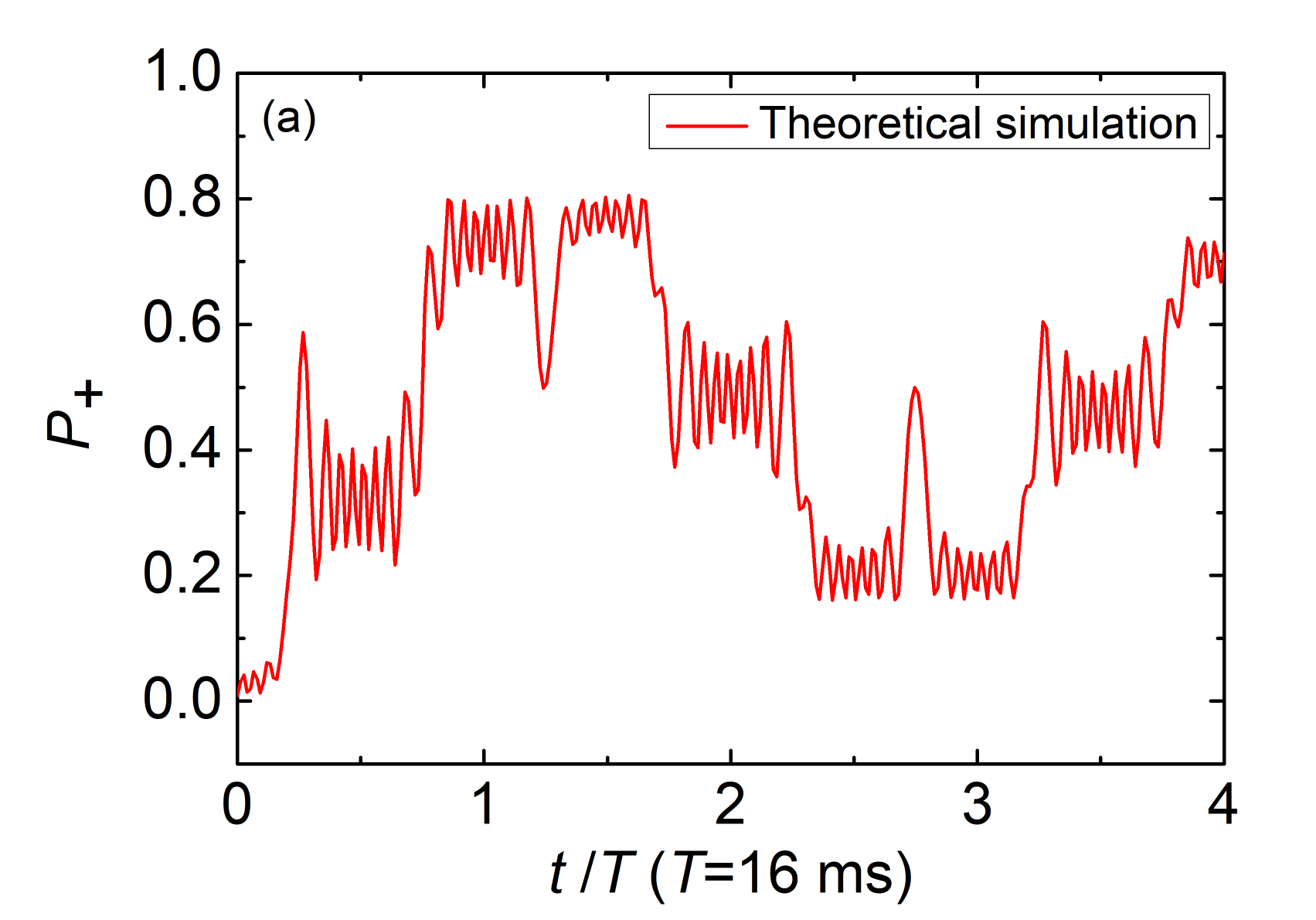}
\includegraphics[width=0.5\textwidth]{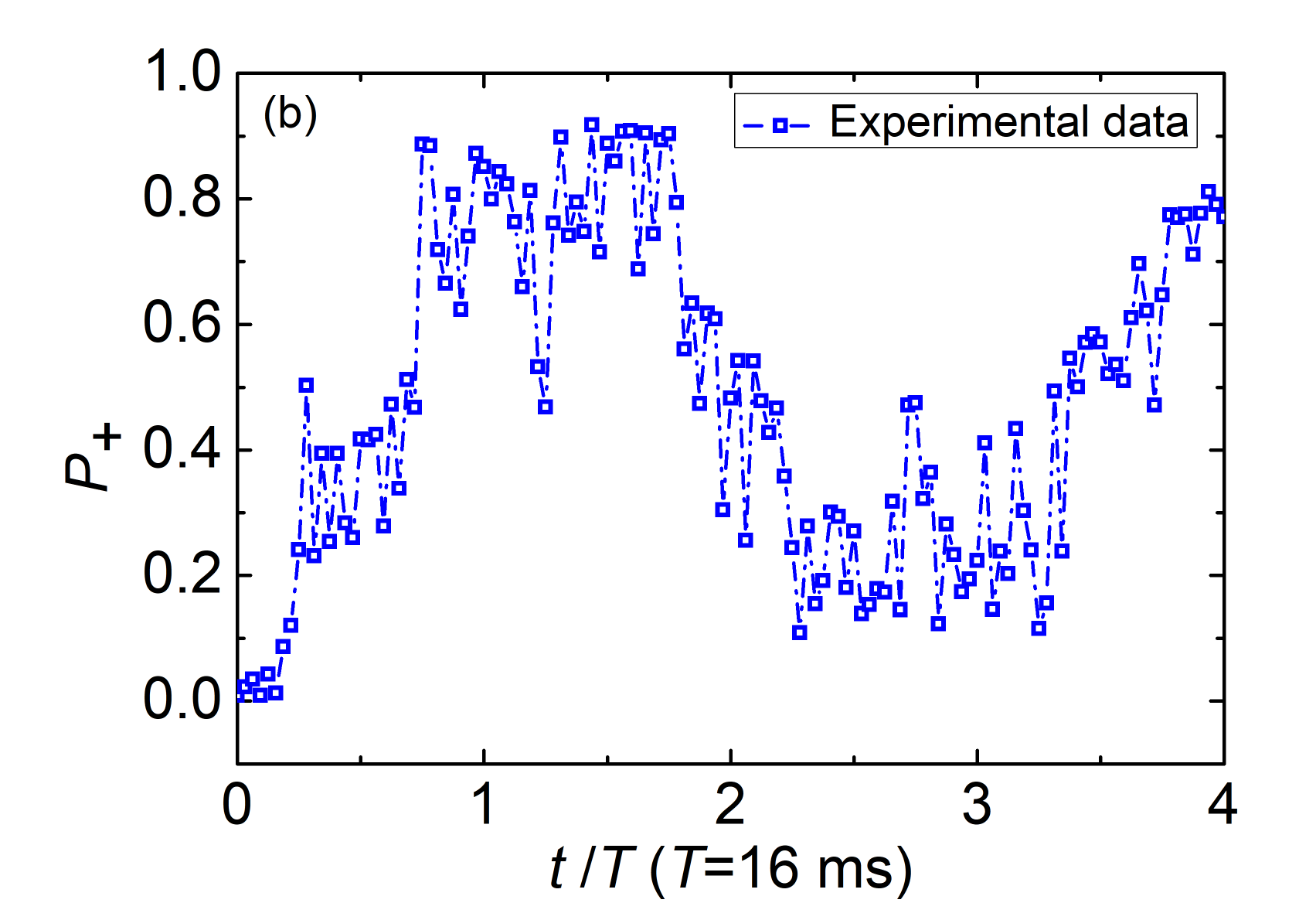}
\caption{Constructive interference in the slow-passage limit over 4 driving periods. The dependence of the excitation rate of the clock transition in the adiabatic basis on the driving period $t$ in (a) theoretical simulation and (b) experimental data.}
\label{slowcon}
\end{figure}
We can see from the experimental results shown in Fig. \ref{fastp}(b) that the constructive interference induced LZROs (upper data) with obvious step-like structures, which agrees well with the numerical simulation Fig. \ref{fastp}(a)(solid red LZRO line). The step-like structure could be interpreted as following:  between two adjacent LZ transitions, atoms adiabatically follow the adiabatic bases, which is very close to the diabatic bases under the experimental parameter. Thus the atom population probability in states $\left| e \right\rangle $ is almost constant between two adjacent LZ transitions. The experimental data in Fig. \ref{fastp} also show that the step structure is very clear within four driving periods but becomes obscure with longer evolution times. The reason for this might be the non-linear drift of the clock laser. Under destructive interference in the fast-passage limit, atoms are frozen in the initial state for each LZ transition, resulting in the coherent destructive tunneling (CDT) phenomenon. This is illustrated in the lower parts of Fig. \ref{fastp}(a) and (b) for the numerical and experimental results, respectively.
\par

\begin{figure}
\includegraphics[width=0.5 \textwidth]{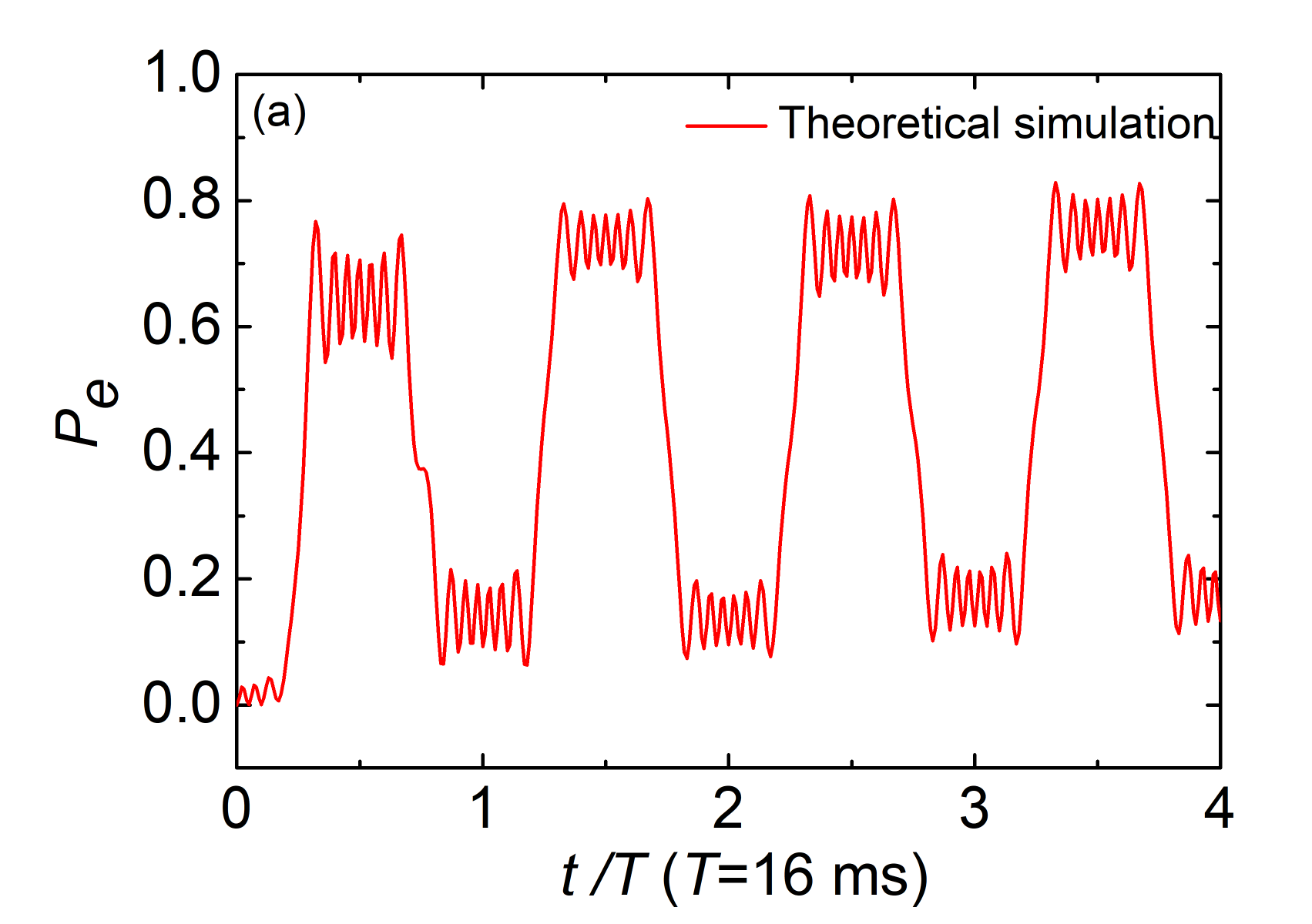}
\includegraphics[width=0.5\textwidth]{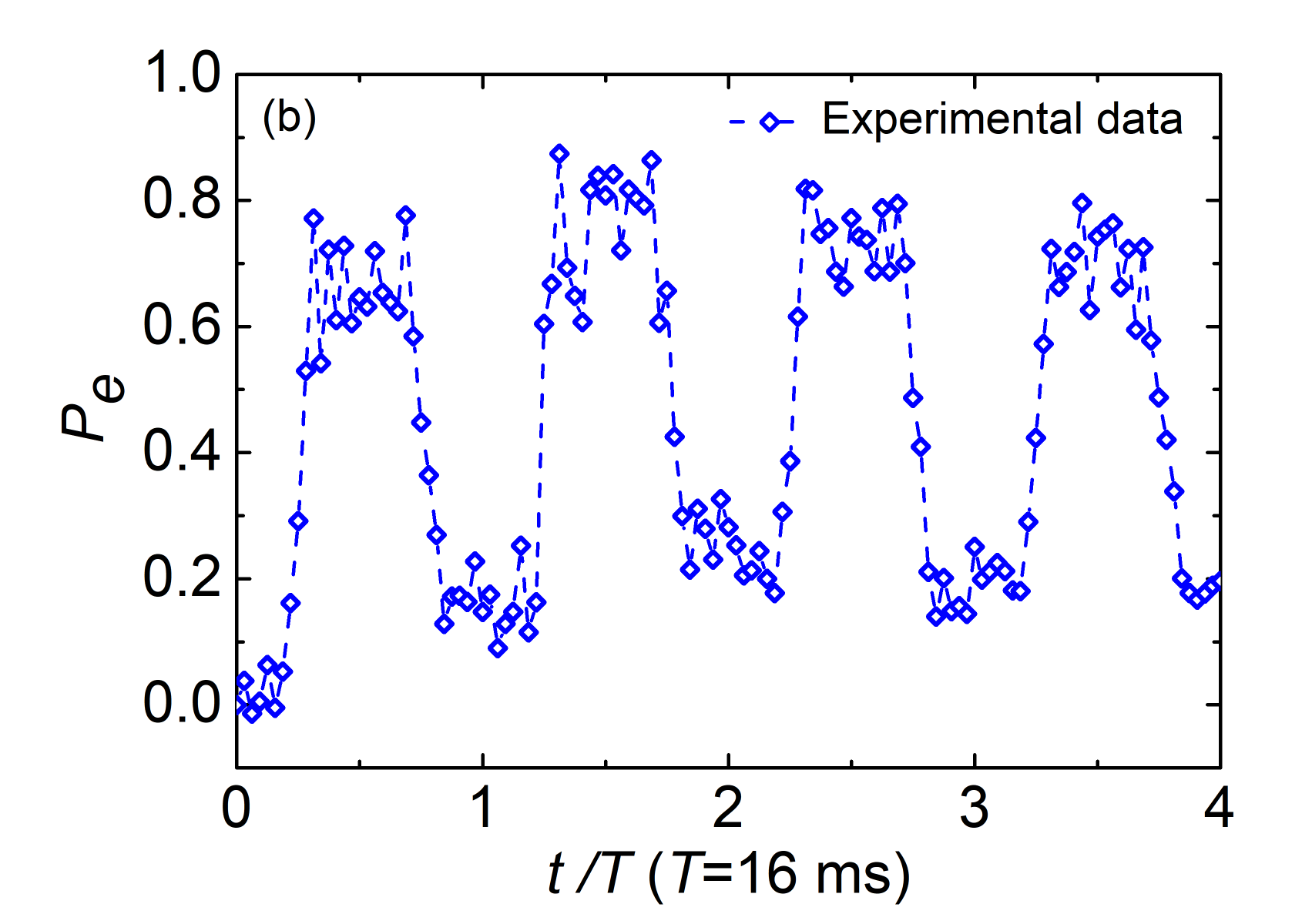}
\caption{Destructive interference at the slow-passage limit. The dependence of excitation rate of the clock transition on the driving period $t$ in diabatic basis under $g$=320 Hz, $A$=20.6. (a)The numerical simulation results. (b) The experimental results.}
%\caption{\label{5}**Comparison of nondriven Rabi oscillations (blue dashed lines) with destructive interference spectra (red line) in slow-passage limit under (a) g=320 Hz, A=20.6 and (b) g=400 Hz, A=21.4, respectively.}
\label{SPLdes}
\end{figure}
\par

\begin{figure}
\includegraphics[width=0.5 \textwidth]{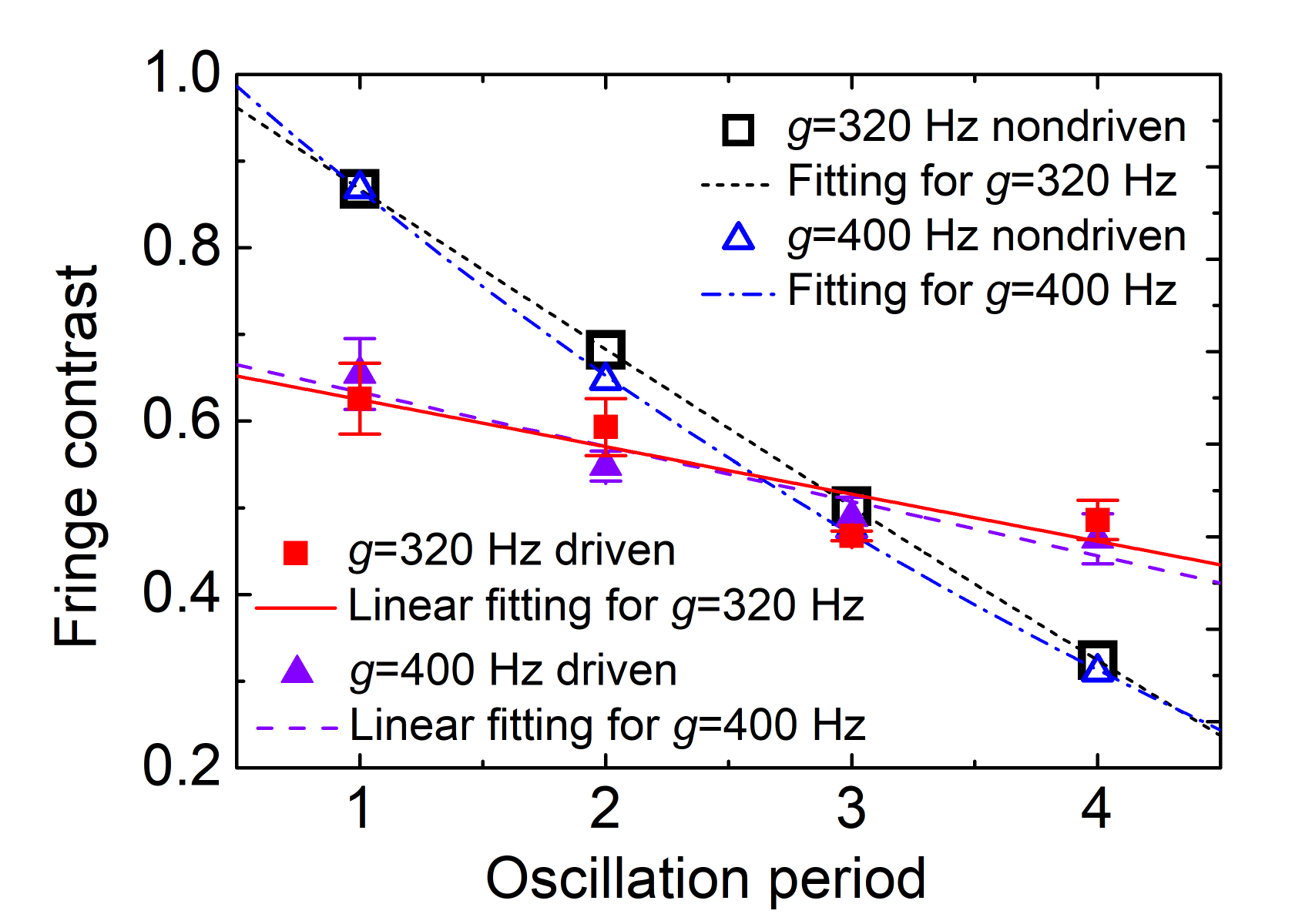}
\caption{Comparison of fringe contrast between nondriven Rabi oscillations (black hollow square with $g$=320 Hz and blue hollow triangle with $g$=400 Hz) and LZ destructive interference spectra (red solid square and solid violet triangle) in slow-passage limit under $g$=320 Hz and $A$=20.6, and $g$=400 Hz and $A$=21.4, respectively. The LZ destructive interference spectra exhibit quite slow decay rate with -0.051 and -0.059 at $g$=320 Hz and $g$=400 Hz, respectively. The exponential fitting ($fringecontrast = offset + D\exp \left( { - {t \mathord{\left/
 {\vphantom {t v}} \right.
 \kern-\nulldelimiterspace} v}} \right)$) for the nondriven fringe contrast show decay constants $v$ of 45.77 for $g$=320 Hz and 6.40 for $g$=400 Hz}
\label{compare}
\end{figure}
Now we turn to the slow-passage limit. In this case, we set $g$ near to $320$ and ${\omega _s}$ to $2\pi\times 62.5$ Hz. Then we measured four driving-periods interference spectrum. The constructive and destructive LZROs were detected under $A=22.2$ and $20.6$, respectively, which give ${P_{LZ}} \approx 0.24$ (slow-passage regime).
\par
For the constructive interference, the oscillation behavior is irregular in the diabatic basis. However, when converting the results to the adiabatic basis, the global interference becomes clearly recognizable. The coarse-grained oscillating behavior in the adiabatic basis shows an explicit LZRO pattern within four driving periods, as shown with in Fig. \ref{slowcon}(b), which is consistent with the numerical simulation presented in Fig. \ref{slowcon}(a). However, after two periods, the fluctuation of each plateau is much larger than predicted by the theoretical simulation, which might be due to the non-linear drift of the clock laser. 

\par
For the destructive interference, because the LZ transition probability is very small, atoms remain in the ground state of the adiabatic basis. When converted to the diabatic basis, the atom population changes dramatically between two plateaus, as shown in Fig. \ref{SPLdes}(a) and (b).The experimental data agree very well with the numerical simulation within four driving periods, except that the oscillation of each plateau is much larger than theoretically predicted. The drastic oscillation between two plateaus can be interpreted as follows: the adiabatic ground state swaps between the diabatic ground and excited states at each level avoided crossing point. 

Moreover, we quantified the dephasing effect under destructive interference conditions within four periods by analyzing the contrast fringe. As shown in Fig. \ref{compare}, the decay of the contrast fringe for two different Rabi coupling strength cases can both be fitted by a slow slope linear function. To compare this with Rabi oscillation, we also measured the contrast fringe within four driving periods under the same experimental conditions, which indicates a fast exponential decay behavior, as shown in Fig. \ref{compare}. One can clearly observe a suppression of dephasing by LZRO. Our system can be considered a Landau-Zener interferometer with uncontrollable noise (non-linear drift); thus, we have demonstrated that LZRO can even suppress dephasing without perfect conditions. The comparison between LZRO and Rabi oscillation for the same measurement time can be found in Appendix B, which provides a more direct indication of the slow decay of LZRO and the fast decay of Rabi oscillation.

\par

\section{Conclusion and outlook}
\label{section4}
In this paper, we observe the LZRO in a $^{87}$Sr optical lattice clock platform using Rabi spectroscopy, by compensating for the linear drift of the clock laser and carefully selecting system parameters. Under fast-passage conditions, the LZRO spectrum for coherent constructive interference manifests as a coarse-grained, step-like oscillation, while coherent destructive interference reveals a CDT effect. In slow-passage conditions, the experimental results for coherent constructive interference also show a coarse-grained, step-like oscillation but are observed in adiabatic rather than diabatic bases. For coherent destructive interference under slow-passage conditions, the atomic population probability oscillates between two widely separated values, each lasting half a period. No decay effect is observed in the spectrum, even with increasing Rabi coupling strength.

Currently, we can only obtain clear LZRO data within 4 driving periods due to the non-linear drift of the clock laser. However, the plateau of the step-like spectrum persists for several milliseconds, which is $10^4$ times longer than in previous NV center experiments \cite{ref27}, thanks to the long coherence time of the optical clock platform. This advantage allows for precise atomic state manipulation, enabling atoms with a broad range of Rabi coupling strengths to approach nearly the same quantum state with a single pulse.

Further work is needed to achieve high-precision quantum state control in atomic ensembles, including reducing non-linear drift effects to extend the number of detectable driving periods and minimizing plateau oscillations.

\begin{acknowledgments}
This work was supported by the Strategic Priority Research Program of the Chinese Academy of Sciences (Grant No. XDB35010202) and the National Natural Science Foundation of China under Grant No. 12274045, No. 12274046, No.12347101 and No. 12203057.
\end{acknowledgments}

\begin{appendix}
\section{Linear drift of clock laser}
\begin{figure}
    \centering
    \includegraphics[width=1\linewidth]{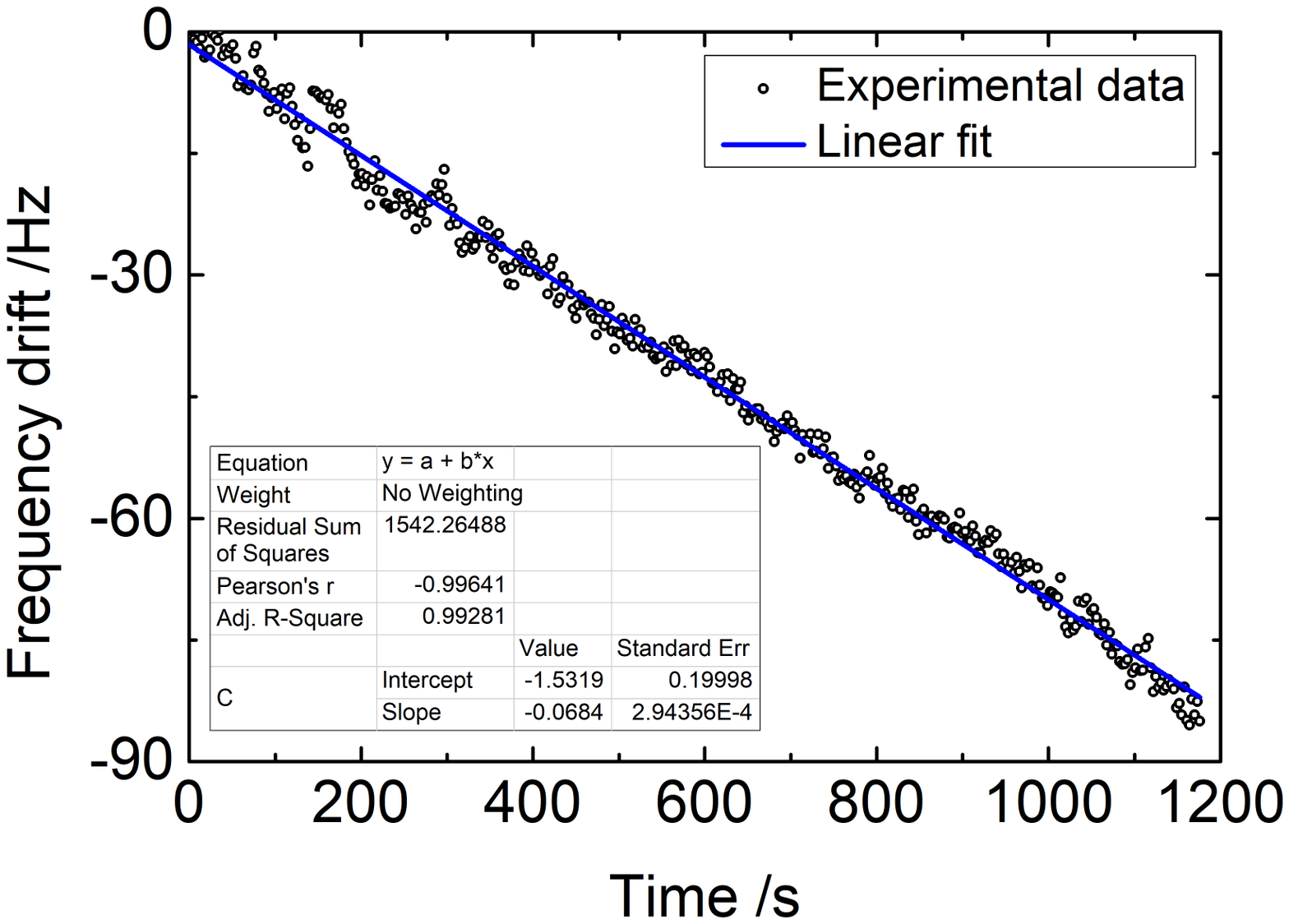}
    \caption{Frequency drift of clock laser. After the clock laser was locked to the transition line, the output frequency of the feedback AOM was recorded, allowing us to measure the frequency drift of the clock laser. This figure shows the frequency variation over a 20-minute period. According to the linear fit, the drift rate is 0.0684 Hz/s. Due to phase noise and feedback loop system noise, the data exhibits short-term frequency fluctuations of ±3 Hz.}
    \label{drift}
\end{figure}

To enable the rapid detection of LZRO, two AOMs are used in the experiment to achieve line drift compensation and frequency modulation of the clock laser, respectively. The sum of the carrier frequencies of the two AOMs equals the difference between the ULE cavity mode frequency and the clock transition frequency. By feedback control in the optical clock system, the linear drift rate of the clock laser varies slowly between 0.06 and 0.09 Hz/s. In Fig. \ref{drift}, we show the frequency drift measured near 20 minutes. Due to daily variations in laboratory temperature, humidity, and other environmental factors, the drift rate of the ULE cavity is slightly different each day. However, the frequency shift over a two-hour period generally follows a linear trend. Therefore, applying a fixed period of linear drift compensation to AOM1 can maintain the clock laser at a fixed-detuning position relative to the clock transition spectrum during the experimental period. The LZRO spectrum can be quickly obtained by gradually increasing the clock laser detection time sequentially. Additionally, scanning the clock laser frequency can facilitate the detection of LZ interferences.
\par

\section{Comparation between the destructive interference LZRO in slow-passage limit and the Rabi oscillation on nondriven condition}

\begin{figure}
    \centering
    \includegraphics[width=1\linewidth]{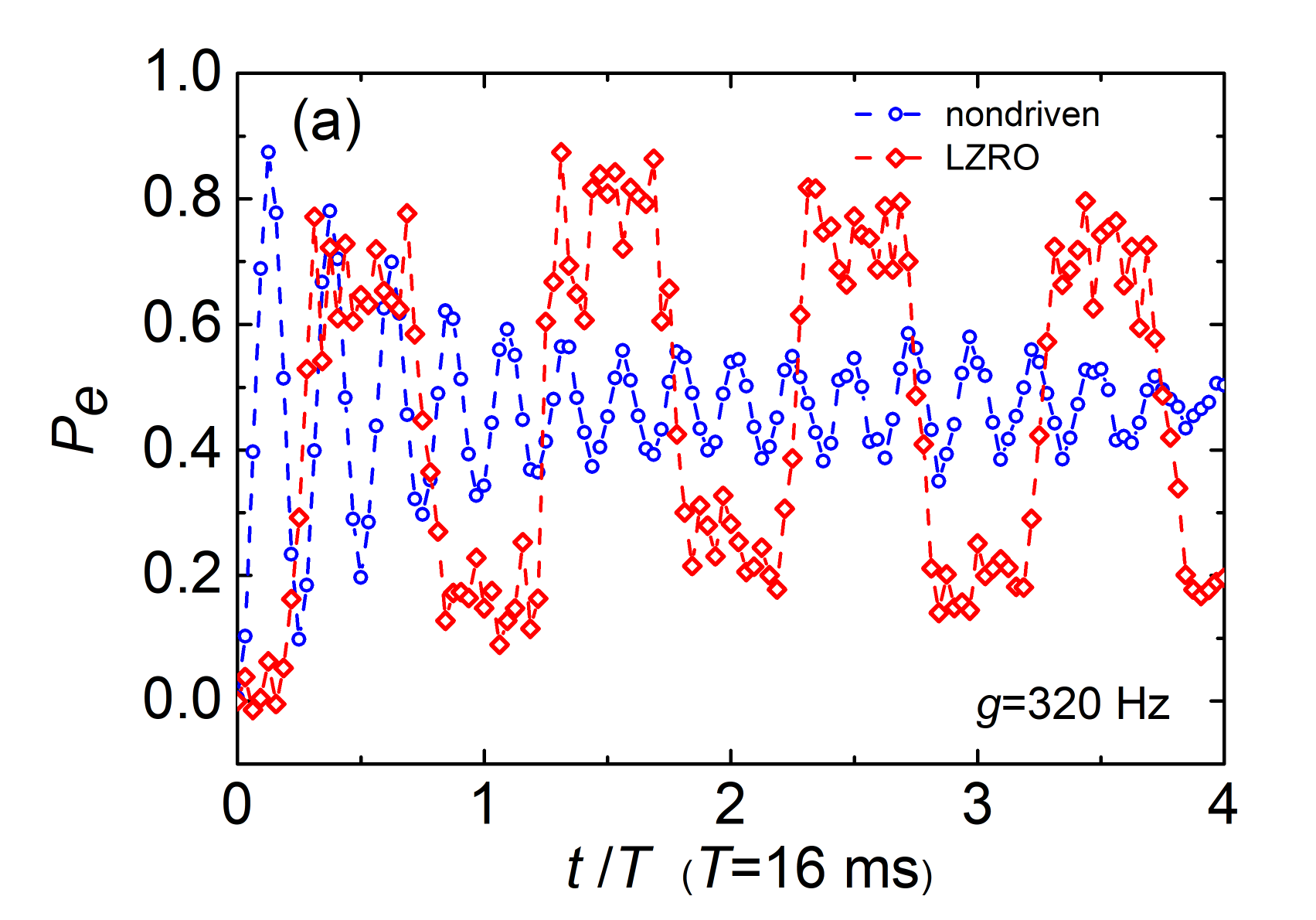}
    \includegraphics[width=1\linewidth]{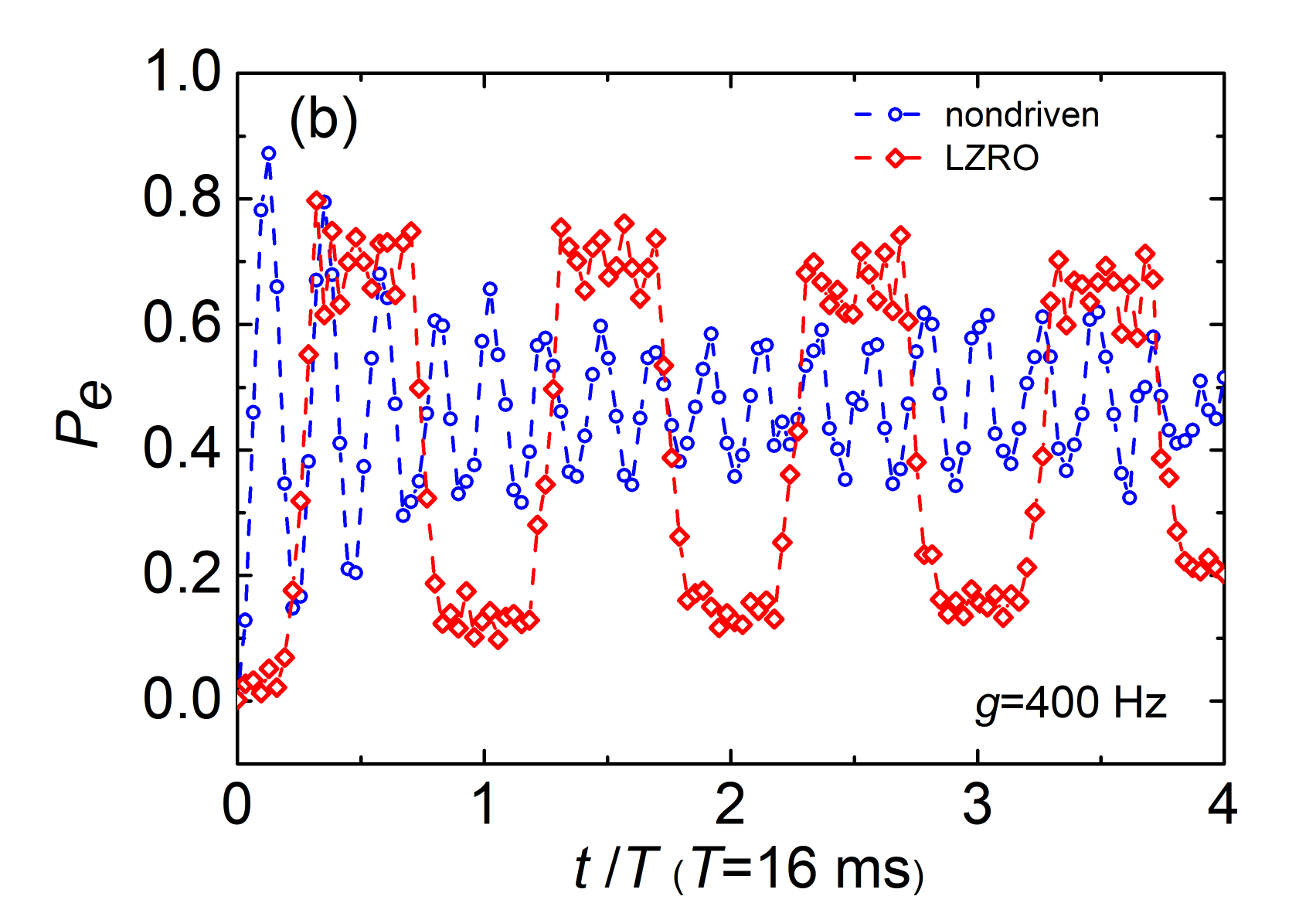}
    \caption{Rabi oscillations of the LZ and nondriven conditions. Diamond plots in red represent LZRO, while circles in blue represent the nondriven Rabi oscillation spectrum. $T$ denotes oscillation period of LZRO. For $g$=320 Hz, $A$=20.6 in (a), and for $g$=400 Hz, $A$=21.4 in (b), respectively.}
    \label{rabicomp}
\end{figure}

To measure more LZRO cycles within an effective coherence time, the driven frequency, ${\omega _s}$, is set to $2\pi\times 62.5$ Hz (with an LZRO period of 16 ms) under the slow passage limit condition. Under LZ destructive conditions, four periods of oscillation are observed, as shown in Fig.\ref{rabicomp}. The nondriven modulation Rabi periods are 1 ms and 8.96 ms at $g$=320 Hz and $g$=400 Hz, respectively.
\par
\FloatBarrier
\end{appendix}
\par

\bibliography{LZSMrefs}

\end{document}